# SARS-CoV-2 mortality in blacks and temperature-sensitivity to an angiotensin-2 receptor blocker

Donald R. Forsdyke,

*Department of Biomedical and Molecular Sciences, Queen's University, Kingston, Ontario, Canada K7L3N6*

*Email address:* forsdyke@queensu.ca

Orcid: 0000-0002-4844-1417

**Pages:** 20.  **Words:** 5739. **Figures:** 3. **References:** 72

## ABSTRACT

Tropical climates provoke adaptations in skin pigmentation and in mechanisms controlling the volume, salt-content and pressure of body fluids. For many whose distant ancestors moved to temperate climes, these adaptations proved harmful: pigmentation decreased by natural selection and susceptibility to hypertension emerged. Now an added risk is lung inflammation from coronavirus that may be furthered by innate immune differences. Hypertension and coronavirus have in common angiotensin converting enzyme 2 (ACE2), which decreases blood pressure and mediates virus entry. In keeping with less detailed studies, a long-term case-report shows that decreased blood pressure induced by blocking a primary angiotensin receptor is supplemented, above critical blocker dosage, by a further temperature-dependent fall, likely mediated by ACE2 and secondary angiotensin receptors. Temperature-dependence suggests a linkage with tropical heritage and an influence of blockers on the progress of coronavirus infections. Positive therapeutic results should result from negation of host pro-inflammatory effects mediated by the primary angiotensin receptor and concomitant promotion of countervailing anti-inflammatory effects mediated by ACE2 through other receptors. These effects may involve innate immune system components (lectin complement pathway, NAD metabolome). Black vulnerability – more likely based on physiological than on socioeconomic differences – provides an important clue that may guide treatments.

*Keywords*: Angiotensin converting enzyme 2, Angiotensin receptor blockers, Glycan shield, Hypertension, Innate immunity, Lectin complement pathway, NAMPT

## 1. Introduction

Humans whose distant ancestors lived in tropical regions can differ biochemically and physiologically from those whose ancestors were in temperate regions. Increased sunlight and heat engender not only evolutionary differences in skin color (Jablonski, 2002), but also in cooling mechanisms that regulate the volume, salt-content and pressure of body fluids (Roman, 1986; Crandall and Wilson, 2015). This has influenced recommendations for treating hypertension, a disease to which those with dark skins are prone (Rostand, 1997; Rostand et al., 2016; Williams et al., 2016). To these major differences in responses to sunlight and heat, the coronavirus (SARS-CoV-2) has added another – an increased mortality of those infected, often from inflammatory complications of lung disease.

Thus, at an early stage of the present epidemic (April 12[th] 2020) the Chicago Department of Public Health reported that, while the percentage mortality in other groups was around 4%, for non-Latino blacks mortality was 22.7%. This extreme difference has been corroborated elsewhere (Kendi, 2020; Li et al., 2020). Since those with pre-existing conditions are particularly vulnerable, the higher mortality could be explained in terms of general ill-health or socioeconomic factors (Burch, 2020). However, given the degree of the disparity, specific differences in underlying physiology could play a role.

As a component of the renin-angiotensin system (RAS) (Jia, 2016), angiotensin II (Ang II) is fundamental to the regulation of blood pressure (BP; Fig. 1). The degrees to which countervailing Ang II receptors are activated could influence, not only BP, but also coronavirus therapy. SARS-CoV-2 enters cells by way of angiotensin converting enzyme 2 (ACE2) (Hoffman et al., 2020), which is generally protective for lung infections (Jia, 2016). Increased *ACE2* gene transcription occurs during SARS-CoV-2 infection (Butler et al., 2020; Vaduganathan et al., 2020; Zhuang et al., 2020; Ziegler et al., 2020). Thus, there is a relationship between coronavirus and a host's physiological mechanisms for the control of the volume, salt-content and pressure of body fluids. Such mechanisms should be most severely tested in hot countries (e.g. sweating with loss of salt and water; Cappuccio and Miller, 2016). It would be expected that, over evolutionary time, just as skin pigmentation became a fixed (non-inducible) adaptation in hot countries, so the fine-tuning of cooling mechanisms by natural selection would have resulted in stable, more extreme, adaptations than in temperate regions.

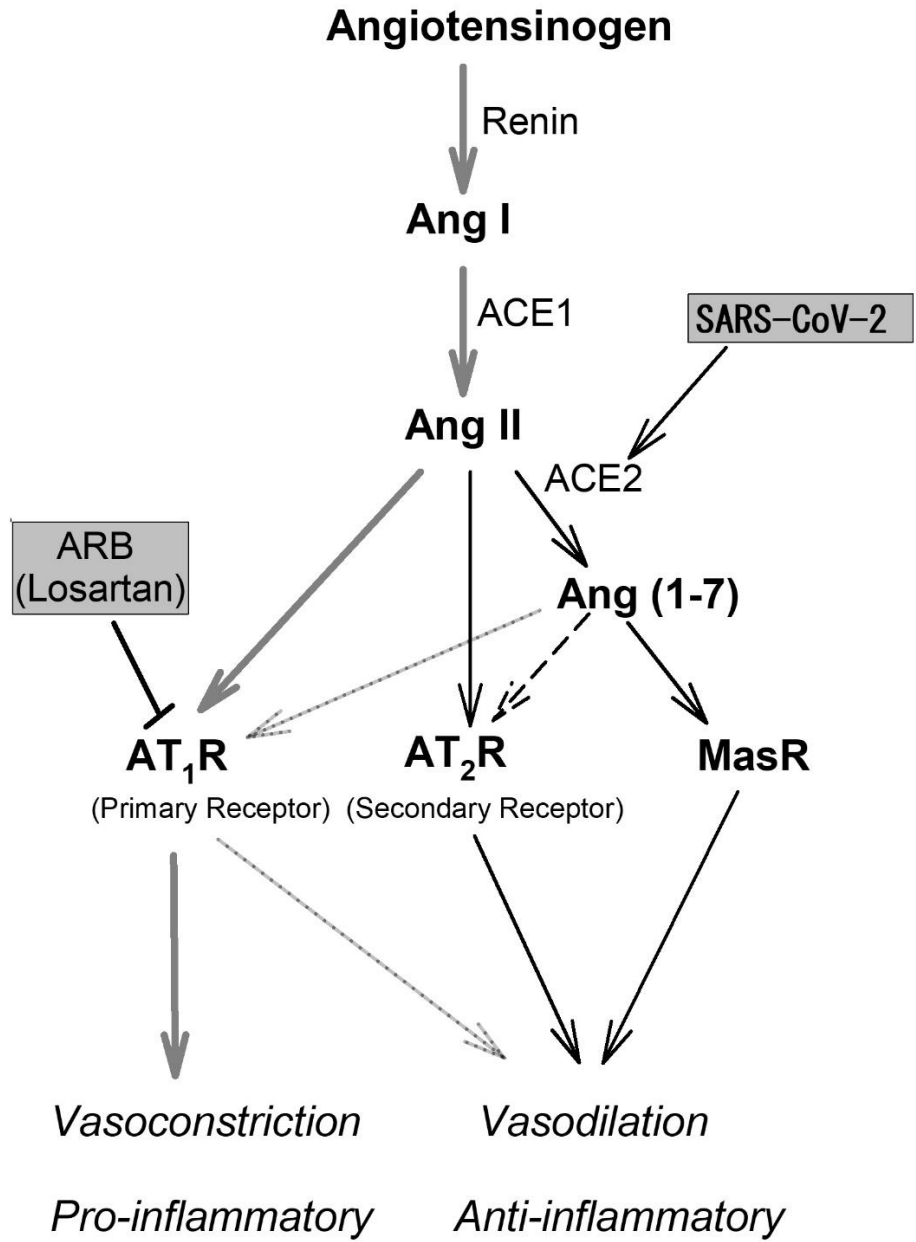

**Fig. 1.** Regulation of BP and inflammation by countervailing receptors in the renin-angiotensin system (RAS). The vasoconstrictory/pro-inflammatory pathway (thick grey arrows) is impeded at the level of the primary Ang II receptor by ARBs. Accumulated Ang II then drives the vasodilatory/anti-inflammatory pathway. Having entered by way of ACE2, SARS-CoV-2 causes lung inflammation that should be opposed by the ARB-induced arrest of pro-inflammatory activity and stimulation of anti-inflammatory activity. The dashed arrow indicates an ARB-dependent activation of the secondary Ang II receptor (Bosnyac et al., 2012). The fine dotted arrow indicates the same result achieved by way of a β-arrestin pathway (Manglik et al., 2020).

However, being advantageous at one location at one point in time is not a guarantee of permanence. Adaptations may further evolve in diaspora. Skin pigmentation proved disadvantageous when populations moved to temperate climes and needed to synthesize vitamin D. Pigmentation was decreased by natural selection, yet it could still be induced to return, albeit at lower levels (tanning), in response to radiation (Jablonski, 2002). Likewise, cooling mechanisms might prove counter-adaptive when challenged, either less severely by temperate climates, or more severely by pathogenic microorganisms. A combination of the two challenges could be fatal.

Knowledge that a RAS component, ACE2, was the SARS-CoV-2 receptor suggested that existing drugs employed to treat hypertension might be repurposed, or modified, to treat coronavirus infections. Indeed, reports from China state that angiotensin receptor blockers (ARBs) can mitigate the severity of SARS-CoV-2 infections (Liu et al., 2020). Their continued in-hospital use is associated with lower mortality risk compared with non-users (Zhang et al. 2020b). My earlier study of a temperature-dependent hypersensitivity of BP to an ARB (losartan), was interpreted in terms of a switch from primary to secondary countervailing receptors (Forsdyke, 2015a). I argue here that this mechanism supports the views, both that ARBs can mitigate the severity of acute inflammatory lung diseases whatever the etiology (Jia, 2016), and that, rather than "racism" (Kendi, 2020), there is a fundamental physiological explanation for the high mortality of those whose distant ancestors lived in tropical regions.

A recent stepwise regression study has identified both African-American origin and environmental temperature as significant variables, and has postulated a marginal vitamin D deficiency to explain the enhanced vulnerability of African-Americans to coronavirus (Li et al. 2020). In contrast, my study, while also implicating temperature as a critical variable, has suggested differential evolved responses to heat-stress that might be influenced by ARBs. This has implications for control of SARS-CoV-2 infections in North America and for meeting the growing threat in Africa.

## 2. Environmental temperature and response to ARBs

Following short-term studies to determine efficacy and minimize adverse side-effects, losartan was introduced in the 1990s for the treatment of hypertension. A case report of its employment by an elderly white male as sole medication over a 12-year period (2003-2014), responded to calls for studies of possible long-term effects (Forsdyke, 2015a). There were also accounts, some anecdotal, of seasonal hyperresponsiveness to antihypertensive medications when environmental temperatures were high (Forsdyke, 2015a; Arakawa et al., 2019*). For the subject of the case report, doses had been adjusted daily with the goal of maintaining home-monitored BP readings (systolic/diastolic) close to 130/80 (mm Hg).

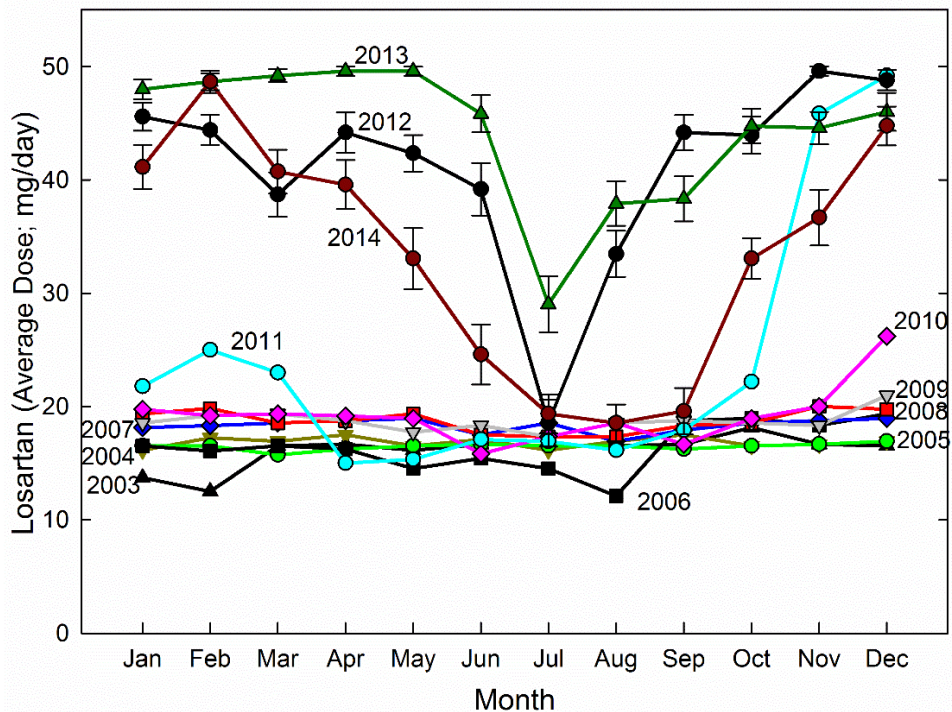

**Fig. 2.** Environment temperature and monthly losartan requirement for a 12-year period (2003–2014). 2003, *black triangles*; 2004, *dark yellow triangles*; 2005, *green circles*; 2006, *black squares*; 2007, *blue diamonds*; 2008, *orange squares*; 2009, *grey triangles*; 2010, *red diamonds*; 2011, *cyan circles*; 2012, *black circles*; 2013, *green diamonds*; 2014, *dark red circles*. Data for 2012–2014 include standard errors. [From Forsdyke 2015a; copyright DRF]

Fig. 2 shows that initially (2003-2010) this goal was consistently achieved with approximately 20 mg losartan daily and no relationship to temperature was evident. Then in late 2011 a need emerged to increase dosage close to 50 mg daily. This had been hinted at by small rises in December 2010 and early 2011. From this higher level, a marked and repeatable summer-time dip emerged (2012-2014). This was clearly seen in plots of dosage against temperature (Fig. 3). Indeed, when extrapolated, the plots indicated that no medication would be needed above 34ºC – a temperature range more consistent with tropical locations than that of the subject (Canada). Since losartan was used world-wide, if not an individual idiosyncrasy then this result had serious implications for the treatment of hypertension in equatorial regions (Imai and Abe, 2013; Sagy et al., 2016).

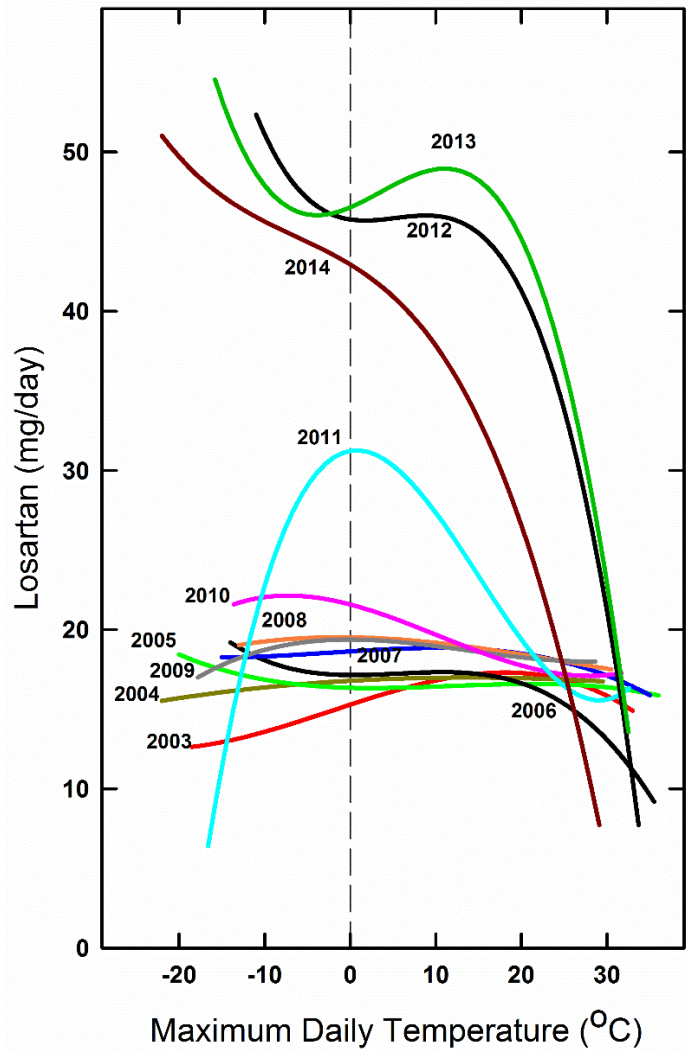

**Fig. 3.** Relationship between daily losartan requirements for the 2003-2014 period and corresponding maximum environmental temperatures. Least-squares regression fits to the points (third order polynomial) are shown as continuous lines for each year. *Line coloring* for different years follows that of Fig. 2. Consecutive $r^2$ values for 2011–2014 were 0.22, 0.38, 0.34, and 0.45. [From Forsdyke 2015a; copyright DRF]

Given that seasonal hyperresponsiveness to anti-hypertensive medications had been the subject – albeit not documented in individual detail – of previous reports, it was hoped that the possibility of individual idiosyncrasy would be discounted by further "crowd sourced" reports (Forsdyke, 2015b). Unlike previous accounts that had not considered detailed mechanisms (e.g. Arakawa et al. 2019; Zhao et al., 2019), it was proposed that the reported phenomena could be explained in terms of differential signaling by countervailing Ang II receptors – inhibition of a *primary* receptor that increased BP (vasoconstriction), and activation, at a critical ARB concentration, of a *secondary* receptor that decreased BP (vasodilation) (Forsdyke, 2015a). The view that these observations are relevant to current considerations of a role of ARBs in the management of SARS-CoV-2 infections was encouraged by rapid developments in the biomedical literature, as will be set out here.

It should be noted that there is generally little incentive for the pharmaceutical industry to carry out sustained long-term studies; hence, the call for "crowd sourcing" (Forsdyke, 2015b). Despite the absence of temperature effects during the first eight years (Figs. 2,3), fortuitously the study was not terminated in 2010. With primary receptors for Ang II (AT$_1$R subtype) blocked by losartan, there should be an increase in the concentration of Ang II that could then suffice to affect the losartan-insensitive, low abundance, secondary receptors (AT$_2$R subtype). Reaction of excess Ang II with the secondary receptors would greatly amplify the fall in BP resulting from the losartan block of primary receptors (Fig. 1). An extensive literature based on both animal and human studies was cited as consistent with this (Forsdyke, 2015a; Bosnyak et al., 2012).

A similar scenario would have the excess of Ang II (an octapeptide) converted by ACE2 to Ang (1-7) (a septapeptide). This truncated Ang II fragment, either by way of another receptor (MasR) (Ferrario et al., 2005; Santos et al., 2018), or by *selectively* influencing the AT$_1$R subtype (Manglik et al., 2020), also lowers BP. Whether acting individually or collectively, these pathways should decrease BP. However, unexplained was the critical dependence on losartan dosage and the relationship to high environmental temperature (Figs. 2, 3). The latter raised the possibility that individuals retaining genetic linkages to ancestors who had adapted to tropical climes, might have particular responses to extrinsic agents, namely to drugs and to the SARS-CoV viruses that bind to ACE2 (Zhang et al., 2020a) – but not to ACE1 (generally referred to as ACE, which has its own specific inhibitors referred to as ACEI).

Of possible relevance to this hypothesis are reports that, as latitude decreases towards the equator, populations tend to have a lower BP (Cabrera et al., 2016), and that sub-Saharan Africans tend to be salt-sensitive regarding increasing BP (Cappuccio and Miller, 2016). Furthermore, while experimental hypertension induced in rats by high salt diets does not respond to ARBs, that induced by the combination of salt and high environmental temperature is responsive (Agbaraolorunpo et al., 2019). Although not concerned with the use of anti-hypertensive medications, a new study by Adam Li and his colleagues of black mortality in SARS-CoV-2 infections (Li et al., 2020), suggests these results are relevant to humans.

## 3.   Role of ARBs in SARS-CoV-2 infection

Given the dependence of SARS-CoV-2 on a host's ACE2 receptor, at issue is whether ARBs can affect, either negatively or positively, the progress of infection (Aronson and Ferner, 2020; Focosi et al., 2020; Guo et al., 2020). Thus, it is proposed that the research community should "better outline the renin-angiotensin system and specifically ACE2 in the pathogenesis of COVID-19," and that clinicians should accumulate more data "to determine if there is a link between the use of ACEIs, ARBs, or both, and COVID-19 mortality and morbidity" (Patel and Verma, 2020). In the absence of such information, various authorities have advised against stopping treatment or switching to a BP treatment that does not involve the renin-angiotensin system (Sommerstein et al., 2020; Quinn et al., 2020). Furthermore, if administered to non-hypertensive patients, ARBs might provoke dangerous degrees of hypotension (Imai and Abe, 2013; Sagy et al., 2016).

Current studies of SARS-CoV-2 are guided by the pioneering studies of SARS-CoV-1 (usually referred to as SARS-CoV), which also enters cells by way of ACE2. SARS-CoV-1 infectivity correlates with ACE2 expression (Hofmann and Pöhlmann, 2004). The Penninger laboratory showed that the lung inflammation responded to anti-hypertensive drugs, with ARBs being particularly effective (Imai et al., 2005; Kuba et al., 2005). So, as recently recollected (Zhang et al., 2020a): "Thus, for SARS-CoV pathogenesis, ACE2 is not only the entry receptor of the virus but also protects from lung injury. We therefore … suggested that in contrast to most other coronaviruses, SARS-CoV became highly lethal because the virus deregulates a lung

protective pathway."

SARS-CoV-2 deaths are mainly associated with respiratory failure resulting from virus-induced inflammation (Jia, 2016; Guo et al., 2020; Li et al., 2017). The ACE2 receptor (like the secondary Ang II receptor, $AT_2R$) has an anti-inflammatory role and is shed or incorporated with the virus on cell entry, so is lost from the cell surface. Sometimes such receptor down-regulation is a viral strategy to prevent superinfection by another, possibly competing, virus (Forsdyke, 2016a). However, this militates against any other roles a receptor might play. While there are indications that SARS-CoV-2 can repress the gene encoding the ACE2 receptor, so that there is no receptor refurbishment (Fadason et al., 2020), detailed transcriptional studies suggest that refurbishment is possible (Butler et al., 2020). Consistent with the human case history discussed here (Forsdyke, 2015a), RAS system-based anti-hypertensive agents (ARBs and ACEI) may stabilize or increase host refurbishment of ACE2 receptors, thus opposing their virus-induced loss (Bastolla, 2020). However, the study of Butler et al. (2020) concluded that with ACEI (*but not with ARBs*) there was "significantly increased risk of intubation and death." Clinical advantages of ARBs and ACEI have been noted by others (Schneeweiss et al., 2020; Zhou et al. 2020).

## 4. Innate Immune differences in African populations

Associations of varying strength have been found between certain RAS genes and susceptibility to hypertension. These associations distinguish African-Americans from other population groups (Zhu et al., 2003). The association approach also supports the case that enhanced susceptibility of African-Americans to SARS-CoV-2 is physiologically based. A recent study of various geographical groups shows a higher variation among Africans of genes encoding proteins that play major roles in innate immunity – e.g. mannose binding lectin 2 (MBL2). Thus, in ancestral Africans natural selection may have favored the explorations of a wider range of responses to environmental challenges than in non-Africans – perhaps bestowing relative disadvantages on the descendent diaspora. Indeed, it is held that the higher variation "might be relevant for the host response to SARS-CoV-2 infection" (Klaassen et al., 2020*). The locations of the amino acid changes in three MBL2 variants would be expected to compromise the ability of these proteins to aggregate into functional units (oligomers), so that decreased activity would be expected. Thus, possibly "variants in genes for proteins involved in the innate immunity add some disadvantage to individuals in combating COVID-19" (Klaassen et al., 2020). Earlier studies of SARS-CoV-1 also concluded that certain MBL variants might be disadvantageous (Ip et al., 2005; Tu et al., 2015).

By virtue of reactivity with mannose-rich surface glycans, MBLs promote phagocytosis of particulate pathogens and affect antibody-independent complement lysis of microbial and mammalian cells. This involves what has become known as the lectin pathway (Forsdyke, 2016b). Thus, as with other innate immune system components, MBLs should assist antiviral host defenses. Indeed, MBLs can react with the "glycan shield" – glycosylated SARS-CoV-1 spike protein (Zhou et al., 2010), and probably also with the oligomannose-rich SARS-CoV-2 spike protein (Chiodo et al., 2020; Gao et al., 2020a).

While these considerations indicate a possible *positive* role of MBLs in combatting SARS-CoV-2, they might also assist cell entry (Ip et al., 2005). Furthermore, proinflammatory cleavage products generated in the complement cascade might enhance SARS-CoV-2 pathogenicity as has been proposed, from mouse studies, they do for SARS-CoV-1 (Gralinski et al., 2018). Indeed, a highly likely *negative* scenario involving the lectin pathway is supported by much emerging evidence *(*Magro et al., 2020). Ting Gao et al. (2020b) have implicated the SARS-CoV-2 nucleocapsid N protein – the most evolutionary flexible of the main coronavirus structural proteins (Dilucca et al., 2020; Gussow et al., 2020) – as activating the complement cascade by way of a lectin pathway intermediate. Studies with the complement inhibitor suramin, which has long been employed to treat African trypanosomiasis, have cast an intriguing light on this.

To be effective against protozoal pathogens, suramin requires a functioning host immune system. It has long been known that the complement-dependent inhibition by lectin of cultured mammalian lymphocytes in calf serum is inhibited at an early stage by suramin (Forsdyke and Milthorp, 1979). From recent experiments with susceptible cells cultured in calf serum it is concluded that "suramin inhibits binding or entry" of SARS-CoV-2 (da Silva et al., 2020). Suramin was found to inhibit if added before, or at the time of, adding virus. However, if added one hour later there was no inhibition. Hence it is suggested that suramin inhibits viral binding to ACE2 receptors and cell entry. Suramin is known to interact with various serum proteins, so there is a possible role for MBLs or other lectin pathway components in the calf serum used for culture; this remains to be explored. These considerations suggest there might be circumstances under which MBLs could both facilitate initial infection (inhibitable by suramin) and enhance pathogenicity (due to inflammatory complement cleavage products). Both of these negative functions might be manifest more in African-Americans by virtue of physiological differences. Since the antimalarial drug chloroquine enhances lectin pathway inhibition (Forsdyke, 1975), it would be contraindicated therapeutically.

African-Americans are often in disadvantaged socioeconomic groups where nutrition may be impaired (Kendi, 2020). Recently, attention has been drawn to a possible role of the enzyme cofactor, nicotinamide adenine dinucleotide ($NAD^+$) in SARS-CoV-2 infection (Heer et al., 2020). Given the wide range of $NAD^+$-dependent metabolic processes (the "NAD metabolome"), it is possible that blacks, by virtue of their distinctive evolution and/or socioeconomic grouping, might be differentially affected. $NAD^+$ levels are normally supplemented by the salvage pathway enzyme, nicotinamide phosphoryl transferase (NAMPT). Greatly increased $NAD^+$ requirements early in immune responses are indicated by enhanced transcription of the corresponding gene (Forsdyke, 2020), which is also found in SARS-CoV-9-infected cells (Heer et al., 2020). Such responses may be promoted by dietary supplements (vitamin B3 derivatives) or by medication with NAMPT activators. When immune responses are deemed disadvantageous, then corresponding NAMPT inhibitors are available (Gerner et al., 2020).

## 5. Conclusions

Investigations on possible ARB therapies are in progress (Patel and Verma, 2020) and are likely to consider the well-reported age and gender differences. What can be added at this time is that the design and interpretation of clinical evaluations should factor in, not only variables such as tropical heritage (e.g. degree of skin pigmentation), but also ambient temperatures, past and present ARB dosage, the stage of the progression of a patient's hypertension, and differences in innate immune defenses of various population groups. In other words, ARBs, by decreasing the pro-inflammatory effects of $AT_1R$ and hence supporting the anti-inflammatory effects of $AT_2R$ and ACE2 receptors, might be remedial under some circumstances, but not others. ARBs would operate, not by way of their ability to lower BP, but by way of their anti-inflammatory activities. Sensitivity to environmental temperature (Forsdyke, 2015a; Li et al., 2020) suggests that patients should be kept in warm environments. Indeed, the case doubling time is longer in tropical regions (Berumen et al., 2020; Notari, 2020), indicating that those with unabridged RAS systems might be helped by higher temperatures (V'kovski et al., 2020; Pirouz et al., 2020).

While much attention is rightly being given to *possibilities* of future effective anti-viral measures (vaccination and drugs selectively affecting viral chemistry), there are pressures to buttress host defenses with what drugs are now available. Thus, "an approach to treating patients with severe COVID-19 infection might be hiding in plain sight;" namely ARBs, perhaps employed in concert with another class of repurposable drug, the statins (Fedson et al., 2020; Zhang et al., 2020c). Others envisage a role for β-arrestins that mediate the desensitization, internalization and ubiquitination of G-protein coupled receptors, such as $AT_1R$. Although it only weakly binds $AT_1Rs$, they point to Ang (1-7) as an "endogenous β-arrestin biased agonist of the $AT_1R$," which is like other arrestins that can be invoked by $AT_1R$ activation (Fig. 1) (Manglik et al., 2020). The repurposing of a drug found to *selectively* modulate $AT_1R$ signaling could activate anti-inflammatory arrestin pathways without the raising of BP or inflammation, which require $AT_1R$ associated G-protein coupled signaling that is controlled by RGS2 (regulator of G-protein signaling 2) (Heximer et al., 2003). If successful, these approaches would allow humoral and cellular herd immunity to develop without concomitant deaths, and hence could decrease requirements for testing and contact-monitoring.

**Acknowledgments and funding**

The arXiv archive has preprint versions of this paper. Queen's University host the author's evolution, immunology and biohistory webpages. This research did not receive any specific grants from funding agencies in the public, commercial or not-for-profit sectors.

# References

Agbaraolorunpo, F.M., Oloyo, A.K., Anigbogu, C.N., Sofola, O.A., 2019. Chronic exposure to high environmental temperature exacerbates sodium retention and worsens the severity of salt-induced hypertension in experimental rats via angiotensin receptor activation. J. Afr. Ass. Physiol. Sci. 7, 109–118.

Arakawa, K., Ibaraki, A., Kawamoto, Y., Tominaga, M., Tsuchihashi, T., 2019. Antihypertensive drug reduction for treated hypertensive patients during the summer. Clin. Exp. Hypertens. 41, 389–393.

Aronson, J.K., Ferner, R.E., 2020. Drugs and the renin-angiotensin system in covid-19. Clinical effects are unpredictable, so treatment decisions must be tailored and pragmatic. Brit. Med. Assoc. J. 369, m1313.

Bastolla, U., 2020.The differential expression of the ACE2 receptor across ages and gender explains the differential lethality of SARS-Cov-2 and suggests possible therapy. arXiv: http://arXiv:2004.07224.

Berumen, J., Schmulson, M., Guerrero, G., Barrera, E., Larriva-Sahd, J., Olaiz, G., et al., 2020. Trends of SARS-Cov-2 infection in 67 countries: Role of climate zone, temperature, humidity and curve behavior of cumulative frequency on duplication time. medRxiv: https://doi.org/10.1101/2020.04.18.20070920.

Bosnyak, S., Widdop, R.E., Denton, K.M., Jones, E.S., 2012. Differential mechanisms of Ang (1-7)-mediated vasodepressor effect in adult and aged candesartan-treated rats. Int. J. Hypertens. 2012, 192567.

Burch, D.S., 2020. Why the virus is a civil rights issue: "The pain will not be shared equally" New York Times, April 21.

Butler, D.J., Mozsary, C., Meydan, C., Danko, D., Foox, J., Rosiene, J., et al., 2020. Host, viral, and environmental transcriptome profiles of the severe acute respiratory syndrome coronavirus 2 (SARS-CoV-2). bioRxiv: https://doi.org/10.1101/2020.04.20.048066.

Cabrera, S.E., Mindell, J.S., Toledo, M., Alvo, M., Ferro, C.J., 2016. Associations of blood pressure with geographical latitude, solar radiation, and ambient temperature: results from the Chilean health survey, 2009-2010. Am. J. Epidemiol. 183, 1071–1073.

Cappuccio, F.P., Miller, M.A., 2016. Cardiovascular disease and hypertension in sub-Saharan Africa: burden, risk and interventions. Intern. Emerg. Med. 11, 299–305.

Chiodo, F., Bruijns, S.C.M., Rodriguez, E., Li, R.J.E., Molinaro, A., Silipo, A., et. al., 2020. Novel ACE2-independent carbohydrate-binding of SARS-CoV-2 spike protein to host lectins and lung microbiota. *bioRxiv:* https://doi.org/10.1101/2020.05.13.092478.

Crandall, C.G., Wilson, T.E., 2015. Human cardiovascular responses to passive heat stress. Compr. Physiol. 5, 17–43.

da Silva, C.S.B., Thaler, M., Tas, A., Ogando, N.S., Bredenbeek, P.J., Ninaber, D.K., et al., 2020. Suramin inhibits SARS-CoV-2 infection in cell culture by interfering with early steps of the replication cycle. *bioRxiv:* https://doi.org/10.1101/2020.05.06.081968.

Dilucca, M., Forcelloni, S., Georgakilas, A.G., Giansanti, A., Pavlopoulou, A., 2020. Codon usage and phenotypic divergences of SARS-CoV-2 Genes. Viruses 12, 498; https://doi.org/10.3390/v12050498.

Fadason, T., Gokuladhas, S., Golovina, E., Ho, D., Farrow, S., Nyaga, D., et al., 2020. A transcription regulatory network within the ACE2 locus may promote a pro-viral environment for SARS-CoV-2 by modulating expression of host factors. bioRxiv: https://doi.org/10.1101/2020.04.14.042002 (2020).

Fedson, D.S., Opal, S.M., Rordam, O., 2020. Hiding in plain sight: an approach to treating patients with severe COVID-19 infection. mBio 11, e00398-20. https://doi.org/10.1128/mBio.00398-20.

Ferrario, C.M., Jessup, J., Chappell, M.C., Averill, D.B., Brosnihan, K. B., Tallant, E.A., et al., 2005. Effect of angiotensin-converting enzyme inhibition and angiotensin II receptor blockers on cardiac angiotensin-converting enzyme 2. Circulation 111, 2605–2610.

Focosi, D., Tuccori, M., Fabrizio, M., 2020. ACE inhibitors and AT1R blockers for COVID-2019: friends or foes? www.preprints.org doi:10.20944/preprints202004.0151.

Forsdyke, D.R., 1975. Evidence for a relationship between chloroquine and complement: Possible implications for the mechanism of action of chloroquine in disease. Can. J. Microbiol. 2l, 1581–1586.

Forsdyke, D.R., Milthorp, P., 1979. Early onset inhibition of lymphocytes in heterologous serum by high concentrations of concanavalin-A: further studies of the role of complement with suramin and heated serum. Int. J. Immunopharmacol, 1,133–139.

Forsdyke, D.R., 2015a. Summertime dosage-dependent hypersensitivity to an angiotensin II receptor blocker. BMC Res. Notes 8, 227. https://doi.org/10.1186/s13104-015-1215-8

Forsdyke, D.R., 2015b. Doctor–scientist–patients who barketh not: the quantified self-movement and crowd-sourcing research. J. Eval. Clin. Pract. 21, 1021–1027.

Forsdyke, D.R., 2016a. Evolutionary Bioinformatics, 3$^{rd}$ ed. Springer, New York. pp. 185-186.

Forsdyke, D.R., 2016b. Almroth Wright, opsonins, innate immunity and the lectin pathway of complement activation: a historical perspective. Microb. Infect. 18, 450–459.

Forsdyke, D.R., 2020. Metabolic optimization of adoptive T-cell transfer cancer immunotherapy: a historical overview. Scand. J. Immunol. (in press) DOI: 10.1111/sji.12929

Gao, C., Zeng, J., Jia, N., Stavenhagen, K., Matsumoto, Y., Zhang, H., et al., 2020a. SARS-CoV-spike protein 1 interacts with multiple innate immune receptors. bioRxiv: https://doi.org/10.1101/2020.07.29.227462

Gao, T., Hu, M., Zhang, X., Li, H., Zhu, L., Liu, H., et al., 2020b. Highly pathogenic coronavirus N protein aggravates lung injury by MASP-2-mediated complement over-activation. medRxiv: https://doi.org/10.1101/2020.03.29.20041962.

Gerner, R.R., Macheiner, S., Reider, S., Siegmund, K., Grabherr, F., Mayr, L., et al., 2020. Targeting NAD immunometabolism limits severe graft-versus-host disease and has potent antileukemic activity. Leukemia 34, 1885-1897.

Gralinski, L.E., Sheahan, T.P., Morrison, T.E., Menachery, V.D., Jensen, K., Leist, S.R., et al., 2018. Complement activation contributes to severe acute respiratory syndrome coronavirus pathogenesis. MBio 9: e01753-18.

Guo, J., Huang, Z., Lin, L., Lv, J., 2020. Coronavirus disease 2019 and cardiovascular disease: a viewpoint on the potential influence of angiotensin-converting enzyme inhibitors/angiotensin receptor blockers on onset and severity of severe acute respiratory syndrome coronavirus 2 infection. J. Am. Heart. Assoc. 9, e016219.

Gussow, A.B., Auslander, N., Faure, G., Wolf, Y.I., Zhang, F., Koonin, E. V., 2020. Genomic determinants of pathogenicity in SARS-CoV-2 and other human coronaviruses. Proc. Natl. Acad. Sci. USA doi/10.1073/pnas.2008176117. [Epub ahead of print]

Heer, C.D., Sanderson, D.J., Voth, L.S., Alhammad, Y.M.O., Schmidt M.S., Trammell, S.A.J., et al., 2020. Coronavirus infection and PARP expression dysregulate the NAD metabolome: an actionable component of innate immunity. bioRxiv: https://doi.org/10.1101/2020.04.17.047480.

Heximer, S.P., Knutsen, R.H., Sun, X., Kaltenbronn, K.M., Rhee, M.H., Peng, N., et al., 2003. Hypertension and prolonged vasoconstrictor signaling in RGS2-deficient mice. J. Clin. Invest. 111, 445–452.

Hofmann, H., Pöhlmann, S., 2004. Cellular entry of the SARS coronavirus. Trends Microbiol. 12, 466–472.

Hoffman, M., Kleine-Weber, H., Pöhlmann, S., 2020. A multibasic cleavage site in the spike protein of SARS-CoV-2 is essential for infection of human lung cells. Mol. Cell 78, 779–784.

Imai, E., Abe, K., 2013. Blood pressure drop in summer may cause acute kidney injury with irreversible reduction on glomerular filtration rate. Clin. Exp. Nephrol. 17, 1–2.

Imai, Y., Kuba, K., Rao, S., Huan, Y., Guo, F., Guan, B., et al., 2005. Angiotensin-converting enzyme 2 protects from severe acute lung failure. Nature 436, 112–116.

Ip, W.K., Chan, K.H., Law, H.K., Tso, G.H., Kong, E.K., Wong, W.H., et al., 2005. Mannose-binding lectin in severe acute respiratory syndrome coronavirus infection. J. Infect. Dis. 191, 1697e704.

Jablonski, N.G., 2012. Living Color. The Biological and Social Meaning of Skin Color. University of California Press, Berkeley.

Jia, H., 2016. Pulmonary angiotensin-converting enzyme 2 (ACE2) and inflammatory lung disease. Shock 46, 239–248.

Kendi, I.X., 2020. Stop blaming black people for dying of the coronavirus. New data from 29 states confirm the extent of the racial disparities. The Atlantic, April 14.

Klaassen, K., Stankovic, B., Zukic, B., Kotur, N., Gasic, V., Pavlovic, S., et al., 2020. Functional prediction and comparative population analysis of variants in genes for proteases and innate immunity related to SARS-CoV-2 infection. bioRxiv: https://doi.org/10.1101/2020.05.13.093690.

Kuba, K., Imai, Y., Rao, S., Gao, H., Guo, F., Guan, B., et al., 2005. A crucial role of angiotensin converting enzyme 2 (ACE2) in SARS coronavirus-induced lung injury. Nature Med. 11, 875–879.

Li, A.Y., Hannah, T.C., Durbin, J., Dreher, N., McAuley, F.M., Marayati, N.F., et al., 2020. Multivariate analysis of factors affecting COVID-19 case and death rate in U.S. counties: the significant effects of black race and temperature. medRxiv: https://doi.org/10.1101/2020.04.17.20069708.

Li, X.C., Zhang, J., Zhuo J.L., 2017. The vasoprotective axes of the renin-angiotensin system: Physiological relevance and therapeutic implications in cardiovascular, hypertensive and kidney diseases. Pharm. Res. 125, 21–38.

Liu, Y., Huang, F., Xu, J., Yang, P., Qin, Y., Cao, M., et al., 2020. Anti-hypertensive angiotensin II receptor blockers associated to mitigation of disease severity in elderly COVID-19 patients. medRxiv: https://doi.org/10.1101/2020.03.20.20039586.

Magro, C., Mulvey, J.J., Berlin, D., Nuovo, G., Salvatore, S., Harp, J., et al., 2020. Complement associated microvascular injury and thrombosis in the pathogenesis of severe COVID-19 infection: A report of five cases. Trans. Res. https://doi.org/10.1016/j.trsl.2020.04.007. [Epub ahead of print]

Manglik, A., Wingler, L.M., Rockman, H.A., Lefkowitz, R.J., 2020. β-Arrestin-biased angiotensin II receptor agonists for COVID-19. Circulation 142, 318–320.

Notari, A., 2020. Temperature dependence of COVID-19 transmission. arXiv: http://arXiv:2003.12417.

Patel, A.B., Verma, A., 2020. COVID-19 and angiotensin-converting enzyme inhibitors and angiotensin receptor blockers. What is the evidence? J. Am. Med. Assoc. doi: 10.1001/jama.2020.4812. [Epub ahead of print]

Pirouz, B., Golmohammadi, A., Masouleh, H.S., Violini, G., Pirouz, B. 2020. Relationship between average daily temperature and average cumulative daily rate of confirmed cases of COVID-19. medRxiv: https://doi.org/10.1101/2020.04.10.20059337.

Quinn, K.L., Fralick, M., Zipursky, J.S., Stall, N.M., 2020. Renin–angiotensin–aldosterone system inhibitors and COVID-19. Can. Med. Assoc. J. 192, E553–554.

Roman, R.J., 1986. Pressure diuresis mechanism in the control of renal function and arterial pressure. Fed. Proc. 45, 2878–2884.

Rostand, S.G., 1997. Ultraviolet light may contribute to geographic and racial blood pressure differences. Hypertension 30, 150–156.

Rostand, S. G., McClure, L.A., Kent, S.T., Judd, S.E., Gutierrez, O.M., 2016. Associations of blood pressure, sunlight, and vitamin D in community-dwelling adults. J. Hypertens. 34, 1704–1710.

Sagy, I., Vodonos, A., Novack, V., Rogachev, B., Haviv, Y.S., Barski, I., 2016. The combined effect of high ambient temperature and antihypertensive treatment on renal function in hospitalized elderly patients. PLoS ONE 11, e0168504.

Santos, R.A.S., Sampaio, W.O., Alzamora, A.C., Motta-Santos, D., Alenina, N., Bader, M., et al., 2018. The ACE2/angiotensin-(1-7)/MAS axis of the renin-angiotensin system: focus on angiotensin-(1-7). Physiol Rev. 98, 505–553.

Schneeweiss, M.C., Leonard, S., Weckstein, A., Schneeweiss, S., Rassen, J.A., 2020. Renin-angiotensin-aldosterone-system inhibitor use in patients with COVID-19 infection and prevention of serious events: a cohort study in commercially insured patients in the US. medRxiv: https://doi.org/10.1101/2020.07.22.20159855.

Sommerstein, R., Kochen, M.M., Messerli, F.H., C. Gräni, C., 2020. Coronavirus disease 2019 (COVID-19): Do angiotensin-converting enzyme inhibitors/angiotensin receptor blockers have a biphasic effect? J. Am. Heart. Assoc. 9, e016509.

Tu, X., Chong, W.P., Zhai, Y., Zhang, H., Zhang, F., Wang, S., et al. 2015. Functional polymorphisms of the CCL2 and MBL genes cumulatively increase susceptibility to severe acute respiratory syndrome coronavirus infection. J. Infect. 71, 101–109.

Vaduganathan, M., Vardeny, O., Michel, T., McMurray, J.J.V., Pfeffer, M.A., Solomon, S.D., 2020. Renin–angiotensin–aldosterone system inhibitors in patients with Covid-19. New Eng. J. Med. 382, 1653–1659.

V'kovski, P., Gultom, M., Steiner, S., Kelly, J., Russeil, J., Bastien Mangeat, B., et al., 2020. Disparate temperature-dependent virus – host dynamics for SARS-CoV-2 and SARS-CoV in the human respiratory epithelium. bioRxiv: https://doi.org/10.1101/2020.04.27.062315.

Williams, S.K., Ravenell, J., Seyedali, S., Nayef, S., Ogedegbe, G., 2016. Hypertension treatment in blacks: discussion of the U.S. clinical practice guidelines. Prog. Cardiovasc. Dis. 59, 282–288.

Zhang, H., Penninger, J.M., Li, Y., Zhong, N., Slutsky A.S., 2020a. Angiotensin-converting enzyme 2 (ACE2) as a SARS-CoV-2 receptor: molecular mechanisms and potential therapeutic target. Intens. Care Med. 46, 586–590.

Zhang, P., Zhu, L., Cai, J., Lei, F., Qin, J.-J., Xie, J., et al., 2020b. Association of inpatient use of angiotensin converting enzyme inhibitors and angiotensin II receptor blockers with mortality among patients with hypertension hospitalized with COVID-19. Circ. Res. 126, 1671–1681.

Zhang, X.-J., Qin, J.-J., Cheng, X., Shen, L., Zhao, Y.-C., Yuan, Y., et al., 2020c. In-hospital use of statins Is associated with a reduced risk of mortality among Individuals with COVID-19. Cell Metab. 32, 176–187.

Zhao, H., Jivraj, S., Moody, A., 2019. 'My blood pressure is low today, do you have the heating on?' The association between indoor temperature and blood pressure. J. Hypertens. 37, 504–512.

Zhou, F., Liu, Y.-M., Xie, J., Li, H., Lei, F., Yang, H., et al., 2020. Comparative impacts of ACE (angiotensin-converting enzyme) inhibitors versus angiotensin II receptor blockers on the risk of COVID-19 mortality. Hypertension 76, e15–e17.

Zhou, Y., Lu, K., Pfefferle, S., Bertram, S., Glowacka, I., Drosten, C., et al., 2010. A single asparagine-linked glycosylation site of the severe acute respiratory syndrome coronavirus spike glycoprotein facilitates inhibition by mannose-binding lectin through multiple mechanisms. J. Virol. 84, 8753–8764.

Zhu, X., Chang, Y.P., Yan, D., Weder, A., Cooper, R., Luke, A., et al., 2003. Associations between hypertension and genes in the renin-angiotensin system. Hypertension 41,1027–1034.

Zhuang, M.-W., Cheng, Y., Zhang, J., Jiang, X.-M., Wang, L., Deng, J., et al., 2020. Increasing host cellular receptor—angiotensin-converting enzyme 2 expression by coronavirus may facilitate 2019-nCoV (or SARS-CoV-2) infection. J. Med. Virol. 2020, 1–9.

Ziegler, C.G.K., Allon, S.J., Nyquist, S.K., Mbano, I.M., Miao, V.N., Tzouanas, C.N., et al., 2020. SARS-CoV-2 receptor ACE2 is an interferon-stimulated gene in human airway epithelial cells and is detected in specific cell subsets across tissues. Cell 181, 1016–1035.